\def \bg{\bigskip} 
\def \no{\noindent} 
 \documentstyle[12pt,aasms4]{article}
\begin{document}

\small

\setcounter{page}{1}
\bg

\bg

\bg

\bg

\begin{center} 
{\large {\bf RADIATION FROM A UNIFORMLY ACCELERATED CHARGE}} 

\end{center}
\bg

\bg
\bg
{\centerline  {\bf Amos Harpaz$^1$  \&  Noam Soker$^1$}} 
\bg

\no 1. Department of Physics, University of Haifa at Oranim, Tivon 36006, ISRAEL
   
\no  phr89ah@vmsa.technion.ac.il 

\no soker@physics.technion.ac.il

\bg
\no {\bf ABSTRACT} 
\bg

The emission of radiation by a uniformly accelerated charge is analyzed.  
 According to the standard approach,  a radiation is observed whenever there 
 is a relative acceleraion between the charge and the observer.  
Analyzing difficulties that arose in the standard approach, we propose that a 
radaition is created whenever  a relative acceleration  between 
the charge  and its own electric field exists.   
The electric field induced by a charge accelerated by an external 
(nongravitational) 
force,  is not accelerated with the charge.   Hence the electric 
field is curved in the 
instantanous rest frame of the accelerated charge.  This 
curvature gives rise to a 
stress force, and the work done to overcome the stress force is 
the source of the 
energy carried by the radiation.  
  In this way, the ``energy balance paradox"  finds 
its solution.  
\bg

\no key words: Principle of Equivalence, Curved Electric Field 

\vfil

\eject

\bg

\no {\bf 1.  Introduction} 
\bg

The emission of radiation from a uniformly accelerated 
charge is considered to be a  well solved problem.  However, 
when the solution of this problem  is treated in its relevance to the principle 
of equivalence, and to observations made by observers located 
in different frames of reference, some contradictions appear to 
exist in the solution.

In the standard spproach to the solution of this problem, 
 which is mostly accepted in the 
physics community,   radiation from a charged particle can be observed 
whenever there is a relative acceleration between the observer 
and the charge (Rohrlich$^{(1)}$ 1965,  see also Rohrlich$^{(2)}$ 1963, Coleman$^{(3)}$ 1961, 
Boulware$^{(4)}$ 1980).   This topic is relevant to the 
principle of equivalence in general relativity,  which   
states the equivalence of a uniformly accelerated system of reference 
in free space to an unaccelerated system, subjected to the action 
of a static homogenous gravitational field (SHGF).  According to 
this principle, no physical observation can lead an observer to 
conclude whether he is located in a uniformly accelerated 
system   in a gravity free space, or if he is at rest 
 in a SHGF.   This principle is  used to analyze the 
possible observations in different  situations, that enable to determine  
whether a radiation can be observed.   Following the analysis 
of Rohrlich$^{(1)}$ (1965), which presents the standard approach in a  wide 
presentation, 
 let us consider a charged particle falling freely in 
SHGF.  According to the principle of equivalence, this particle is 
in  an inertial system.   In order to calculate the radiation from 
this particle, one should determine the electromagnetic fields of the 
charged particle.   

According to Rohrlich$^{(1)}$ (1965), there are three criteria for the existence of 
 a radiation  field:  

(1) The source of the electric  field  is accelerating. 

In a case that we only know the velocities and not the acceleration, we can use a   
second criterion:   

(2)  The field falls  with the distance as  $1/r$. 

However, the use of this criterion is convenient  in the wave zone, where the 
distance from the source is larger than the wave length ($ r \gg \lambda$).  
 
There is still another local criterion. 
      
(3) For this criterion, one integrates the electromagnetic energy tensor over a sphere 
of radius $r$ that circumference the source,  at time $t = t_0 + r/c$, where the source was 
located at the center of the sphere at $ t = t_0$.    The vanishing of this integral shows that 
no radiation is emitted.  The importance of this criterion is that it can be performed at 
any (short) distance from the source.      

In \S 2 we analyze the standard approach and point to several weak points 
in this approach.   In \S 3 we present an alternative approach, 
according to which 
radiation is created due to stress force that exists in the electric field of 
the accelerated charge,  and in \S 4 the energy carried by the  radiation 
is calculated, using the work done to overcome the stress force of the field.   
  
\no {\bf 2.  Analysis of the Standard Approach} 
\bg

Using Maxwell's equations, we can calculate   
electromagnetic fields  in inertial frames of reference (in flat 
space-time, using the tools of the special theory of 
relativity).  These fields are then transformed to accelerated frames of 
reference (or to curved space-time by using the 
transformations of the general theory), and by these transformations 
the fields in the non-inertial systems can be found.   In order  to 
analyze the standard  approach,   
consider an observer falling freely in an SHGF.   A freely falling system 
of reference is    
an inertial system of reference for him.  A charged particle falling freely 
parallel to the observer  in 
the same SHGF is static relative to the freely falling observer.  
The observer will observe a static unperturbed  culoumb field, and no 
radiation 
from the freely falling charge.  A neutral particle, having an 
equal mass,  falling freely in 
parallel to the charged particle, will fall exactly with the 
same acceleration and velocity 
as the charged particle, because both particles lose no energy by radiation, 
and the work done on the particles by the gravitational field that  creates  
the kinetic energy of the particles is equal for both of them.    
  The observer finds no difference 
between the response of the two particles to the SHGF.   This observation 
supports the principle of equivalence, because the observer cannot distinguish 
between the situation in which he and the particles are falling freely in 
SHGF, or if all of them are located in a static gravity-free system 
of reference (Rohrlich$^{(1)}$ 1965).  

An observer located in a static laboratory in the gravitational field, 
will observe that the two particles are uniformly accelerated in the SHGF.  
In order to find the electromagnetic fields in his lab, he performs a 
transformation of the fields of the charged particle  
 calculated in the inertial frame,  to his 
system of reference.  This transformation creates a magnetic field in the 
lab system, in addition to the change in the electric field caused by the 
transformation.  According to the standard   
approach, the  
 existence of these fields in the lab system gives rise 
to a radiation field, and  
the observer in the lab will observe this 
radiation, and conclude that a freely falling charge in SHGF 
 radiates (Rohrlich$^{(1)}$ 1965).  
Certainly, the neutral particle falling parallel to the charged particle 
emits no radiation.  

However, this picture of the standard approach includes several  weak 
points:  

(1)  The lab observer can perform transformation of the velocities 
of the two particles (the charged particle and the neutral one), to 
his system of reference.   
We recall that the velocities of the two particles as observed by 
 the freely-falling 
observer are equal, and the 
transformations of the velocities are of pure kinematic 
 nature.     If the two velocities were equal in the 
inertial system, they will also be equal in the lab system of 
reference, although one of the particles is radiating in the 
lab system.   
    The question now arises, what is the source 
of the energy carried by the radiation from the charged particle?  
 
 (2) Another  question arises: is the radiation emitted 
 by a  source an objective 
phenomenon, or is it a relative phenomenon, that exists for  one 
observer, and does not exist for another?     
This question may be put more strongly:  Suppose that the lab observer that 
observes the radiation, finds that the radiation is absorbed by  another 
object, and causes some physical process in that object, 
like a transition  to an excited 
energy level.   Would the freely falling observer observe this process?  
What will be his explanation for the energy source that created this 
physical process?  

These questions reveal a contradiction in the standard approach.   
 The existence of radiation from 
a certain source is a universal phenomenon that can be observed, 
directly or indirectly, by any observer, without regard to his state 
of motion.   We shall  show later that the relative acceleration 
considered by Rohrlich as the  condition for the observation of radiation, 
is actually  the relative acceleration between the accelerated charge and its  
own electric field, and not between the charge and the  observer.  

(3) Another  problem concerned with uniform acceleration is 
the vanishing of the radiation reaction force  (see also 
Singal$^{(5)}$  1997).   

The Dirac-Lorentz equation for a charged particle is:  

$$ m a^{\mu} = F_{in}^{\mu} + F_{ext}^{\mu} + \Gamma^{\mu} , \eqno(1)  $$  

\no where $m$ is the particle mass, $a^{\mu}$ is the 4-vector  acceleration,   
  $F_{in}^{\mu}$ is the incoming radiation field,  $F_{ext}^{\mu}$ is 
the external force that drives the particle, and  $\Gamma^{\mu}$ is Abraham 
four-vector of the radiation reaction: 
$$ \Gamma^{\mu}  =  {2 e^2 \over 3 c^3}\left( \dot{a}^{\mu} - 
{1 \over c^2} a^{\lambda} a_{\lambda} v^{\mu} \right) ,\eqno(2) $$ 
\no  where $v^{\mu}$ is the 4-vector velocity of the source: $v^{\mu} = 
 \left(\matrix  {\gamma c \cr  
\gamma \vec{v}}\right)$.

$\Gamma$ was considered the ``radiation-reaction force", and the 
work done against this force, was considered as the source  
for the energy carried by the radiation.  
  The first term in $\Gamma$, ${2 e^2 \over 3 c^3} \dot{a}^{\mu} $,   
 is called the ``Schott" term.  
For a uniformly accelerated particle, the motion is a hyperbolic motion 
(Rindler$^{(6)}$  1966), $ \dot{a}^{\mu} = {1 \over c^2} a^{\lambda}a_{\lambda} v^{\mu}$, 
and  $\Gamma^{\mu}$  vanishes.  

The   problem of the vanishing of the radiation reaction force   is called  
 the ``energy balance paradox", where the 
vanishing of this   force in such a motion  gives 
rise to the question, what is the source of the energy carried away  
 by the radiation. 
Several solutions were suggested for this paradox.  Let us mention here 
the one suggested by Leibovitz and Peres$^{(7)}$ (1963).  
 This solution assumes 
the existence of an infinite charged plane, whose total charge is equal and 
opposite in sign to the charge of the accelerated particle, and it recesses   
with the speed of light in the opposite direction to that of the acceleration.  
The interaction between this plane and the accelerated charge is the source of 
the energy carried by the radiation.  This solution is far from being 
satisfactory.    

  The solution suggested by the standard 
approach (Rohrolich$^{(1)}$ 1965) assumes that    
 the Schott term, which is a part of 
 ``Abraham four-vector" is
somehow isolated from the radiation reaction force, and is adaptted to the acceleration 
term in the equation of motion.   
According to this approach, the radiation reaction force is only:  ${-2 e^2 \over 3 c^3} {1 \over c^2} 
a^{\lambda} a_{\lambda} v^{\mu}$.    
  To solve the equation of motion in its new 
form, a multiplication by an integration factor  ($e^{-\tau/\tau_0}$)  
 is needed.  This term, that includes the proper time in an exponential form, leads later 
to divergent solutions that must be discarded.  
The inclusion of the Schott term with the acceleration brings up another 
difficulty: the acceleration is influenced by a force that will act 
a short time later.   This ``backward action" in time (preacceleration),  violates 
causality.   Rohrlich$^{(1)}$ (1965) argues that this  backward action 
takes place over time scale that cannot be measured experimentally.  
   However, if the Schott term is left with the ``force" 
term, it is cancelled out in the case of uniform motion, by 
the second term in   $\Gamma^{\mu}$,  and the above mentioned difficulties  
are avoided.  
  
 (4) Another difficulty concerned with the standard approach  
(see also Singal $^{(5)}$ 1997),  is the 
fact that for a 
zero velocity, the radiation from the uniformly accelerated 
charge has a mirror symmetry 
with respect to a plane located at the charge,  perpendicular 
to the direction of the acceleration (see Jackson$^{(8)}$ 1975, 
eq. 14.39).   Due to this symmetry, the radiation imparts no momentum 
to the radiating charge, 
and there is indeed, no reaction force of the radiation on the 
radiating charge.  
    This consideration justifies the vanishing of 
$ \Gamma^{\mu}$ (which represents the radiation reaction force), 
and the removal of the 
Schott  term  from  this expression leaves us with a radiation 
 reaction force which 
physically does not exist.

The solution to the energy balance paradox is found in  
  an alternative approach, proposed in the 
present work. We find that the electric field of an accelerated charge is 
curved relative to  the  instantanous system of reference of the charge, 
and the interaction between the charge and its own curved electric field  
gives rise to the radiation.  
  In this solution the assumption about the Schott term 
 is not necessary and the difficulties that emerge from the use 
of this assumption are avoided, nor an assumption of an infinite charged plane 
is needed (Leibovitz and Peres$^{(7)}$ 1963).   The source of the 
radiated energy is found in the interaction between the accelerated 
charge and its own curved electric field.  

Recently, two new papers that deal with the question of
accelerated charge appeared in the 
literature  (Singal $^{(5)}$ 1997,  Parrot $^{(9)}$  1997).   Singal,
after criticizing the standard approach, concludes that a
uniformly accelerated charge does not radiate.  This conclusion
resembles that of Rosen $^{(10)}$ (1962), who used similar arguements
to conclude that a freely falling charge does not radiate.
We do not agree with the conclusion of Singal, but we agree with
the majority of his criticizm of the standard approach.  
  Singal's points  (i), (ii), (iii)  (page 1372),
raise  the question of the radiation reaction force, and indeed
point at inconsistencies in the standard approach.  We agree
with this criticizm, and we believe that our conjencture about the
interaction of the accelerated charge with its own curved electric
field, supplies the correct answer to this problem.
Singal's point  (iv)  is that the motion of a uniformly accelerated
charge does not contain any intrinsic time scale, that can determine
a characteristic frequency for the radiation.  We claim that
there is an intrinsic time scale, defined similarily to the
definition given by Jackson $^{(8)}$ (1974, pp 667).  Following his
approach, we define a characteristic frequency $\omega_c$   as:  
$\omega_c = c/R_c \sim a/c $, where  
   $R_c$ is the radius of curvature of the    
electric field.   This radius of curvature plays a crucial role
in the creation of the stress force in the curved field (see eq. 4), 
 which gives
rise to the radiation.
Parrot $^{(9)}$ (1997)  does not accept the conclusion of Singal and
discusses this topic.  Actually, he responds to an earlier paper
of Singal $^{(11)}$ (1995), and only partly to the present paper of
Singal.  He claimes that the fault in Singal's approach is
in that he treats an "acceleration for all times".  Parrot claims
that such a treatment is not physical, and he also presents  
 examples in which
such a treatment leads to inconsistencies.  He comes to 
the conclusion that a uniformly accelerated charge for an 
arbitrary long (but finite) time, does  radiate.
However, our approach is entirely different, and we do not
lean on Parrot's arguements in our conjencture.

\bg

\no {\bf 3.  The Problem} 
\bg

According to classical electrodynamics the power radiated by an  accelerated 
charged particle  is  (Jackson$^{(8)}$ 1974)   
$$ P = {2 \over 3} {e^2 a^2 \over c^3} , \eqno(3)   $$ 

\no where $e$ is the particle charge,  $a$ is its acceleration, and 
$c$ is the speed of light.   
The existence  of a radiation depends on the existence of an acceleration.  
 In the frame of special relativity,  
 nonaccelerated charged particles do not radiate.   Therefore,  
 the existence of radiation provides an absolute measure to distinguish 
between accelerated charged particles and nonaccelerated 
 ones (Rohrlich$^{(1)}$ 1965).  
The question is: when one speaks about acceleration, to what system this 
acceleration is related.  In the standard approach, the acceleration  is 
measured relative    
 to the observer, and if the observer and the charge are accelerated 
with the same acceleration, their relative acceleration vanishes, and such an 
observer will not detect radiation, the same as an observer at rest when 
observing a static charge.  

We argue that the calculations presented by Rohrlich$^{(1)}$ (1965) 
 for this case are 
correct, but 
we present a different physical interpretation to these calculations.  
   The relative acceleration 
that matters in the present case is the 
 {\it relative acceleration between the charge 
and its electric field.}      

Let us elaborate on this example in more detail: 
The electric field is defined as a property of space (around the electric 
charge).   The field at each point (x)  tells us what force will act on 
a unit charge when located at this point.  The field is induced on space 
by a charge.  Once it is induced, it is an independent physical entity,  
and its behaviour is determined by the nature of space on which it is 
induced.  When an electric charge is accelerated by an external 
(nongravitational) force,  the space surrounding the charge is not affected 
by this force.  The electric field created by the charge at each moment, 
is attributed to space, and its behaviour is determined by the space.   Hence, 
the field is not accelerated by the external force, and the charge is 
accelerated relative to its field.  This situation is the cause for the 
creation of the radiation field   
(Jackson$^{(8)}$  1974, eq. 14.14)          


 Without stating it explicitly, Jackson considers an acceleration
relative to an inertial   frame of reference that can be describedd 
by  special relativity.  
 Ordinarily, when general relativity is considered, the inertial   
frame of reference should be replaced by a freely falling frame of 
reference characterized by a set of  geodesics that cover this system.     
Acceleration now is related to the local system   of geodesics.  
The electric field that was induced on space by the electric charge, 
follows the local sysmtem  of geodesics, while the charged particle 
accelerates relative to this frame of reference.  There exists 
a relative acceleration between the electric field that follows  
the system of reference defined by the    
the geodesics, and the accelerated charge that does not follow this system.     

A freely falling charge   
  in a uniform gravitational 
field,  moves along a geodesic line, and it is not subjected to any 
external force.   
 Any physical entity located in the same space is subjected to the 
same   gravitational field, and will follow  similar   geodesics.   
 The charge and the field created by this charge are located in the same 
frame of reference; and in that frame, their relative situation is similar 
to the one existing between a static charge and its field in 
a free space (Singal$^{(11)}$ 1995). 
 We conclude, in accordance with Einstein principle of equivalence, 
that a freely-falling charge does not radiate.  

This charge can be observed by an observer falling freely parallel to 
the charge.   Certainly, this observer will not detect radiation.  The standard 
approach came to the same conclusion (Rohrlich$^{(1)}$ 1965).   
 When a static observer, 
sitted in his lab at rest in the gravitational field, observes the 
 same freely falling 
charge,  the physical situation 
of the  charge does not change.   We conclude 
that this observer finds that the charge and its 
electric field have the same acceleration and  no radiation 
can be emitted from the charge.   The lab observer will come to the conclusion 
that  a freely falling charge does not radiate.  This is opposite to the 
 conclusion of the standard approach.

We describe the electric field of a single charge as field lines emanating 
from the charge to infinity.   
Any change in the field, induced by the charge on the space surrounding it,   
proceeds with a finite velocity, $c$.  
The field lines of a charge, are related  to the system of  
 geodesics, and for a static charge 
in free space (where spherical symmetry holds), they are straight lines.  
 The field lines are influenced by the situation of the charge at the 
moment they ``leave" the charge.   Once they left the charge, they 
 are related to  the local system of geodesics that characterizes the space,  
and are not influenced any more by the further motion of the charge.  
Hence, when a charge is accelerated in a free space, the local 
system of geodesics   to which the field lines are related,   forms a curved system 
 relative to  the instantaneous system of the charge.  The lines of the 
electric field of the charge  are  curved lines in the system of the   
(accelerated) charge that created the field (see figure 2 
in Singal $^{(5)}$ 1997).   This curvature of the 
electric field  gives 
rise to a   stress force between the charge and its electric field, 
  and the action  of 
 this force creates the radiation.


\bg
\no {\bf 4. The Energy Source of the Radiation } 
\bg

 Let us study quantitatively 
the work done by the stress force of the curved  electric field.
 The radiation emitted by a charged particle moving at a constant 
acceleration, $a$, relative to the instantaneous rest frame of the 
charge  has a cylindrical symmetry around the direction of acceleration. 
When we start with a zero velocity, there is  a mirror symmetry with 
respect to a plane perpendicular 
to the acceleration.  
 Therefore the radiation imparts no momentum to the radiating charge, 
and the force, $F_{\rm acc}$, that creates the acceleration $a$ is the 
same as the force required to accelerate a neutral particle with 
the same mass and acceleration at zero velocity.   The work done by  this force 
 creates the kinetic energy of the accelerated particle.   
For the creation of the radiation, an additional work  is required.  
What  force performs this work?

The force that creates the radiation is the force needed to 
overcome  the stress force of the 
electric field of the particle, which is curved in the instantaneous 
rest frame  of the particle.    This extra force, has to be 
supplied by the external force,   
 in addition to the force needed to   create the 
kinetic energy of the accelerated object.  
 Therefore the total  work performed by the external force 
   can be decomposed into two parts:  one 
part creates the kinetic energy of the particle, and the 
second part creates the  energy carried by the radiation. 
The energy balance paradox is thus solved, where  
the energy done in overcoming  
 the stress force of the curved electric 
field, creates the energy carried by the radiation, although the radiation 
reaction force, $\Gamma^{\mu}$, vanishes for uniformly accelerated charge.   
We shall sum over the stress force of the field, $f_E$, and 
calculate the work done against this force.

In order to sum over $f_E$,   we have to 
integrate over a sphere whose center is located on the charge.  
Naturally, such an integration involves a divergence (at the center). 
To avoid such a divergence, we take as the lower limit of the 
integration a small distance from the center, $r = c \Delta t$, (where 
$\Delta t$ is infinitsimal), and later we demand that 
$\Delta t  \rightarrow 0$.  
  We calculate the work done by the stress 
force in the volume defined by  $c \Delta t < r < r_{up}$, 
where  $c^2/a \gg r_{up} \gg c \Delta t$.   
These calculations are performed in a system of reference $S$, 
defined by the geodesics, and momentarily coincides with the frame of 
reference of the accelerated charge at the charge location, 
at time $t = 0$.   

 The force per unit volume  due to the electric stress is 
$f_E = E^2/(4 \pi R_c)$, where $E$  is the electric field, and 
$R_c$ is the radius of curvature of the   
field lines.
 The radius of curvature is calculated by using the formulae for 
 hyperbolic
motion  (Rindler$^{(5)}$ 1966).   
 It can be easily shown that in the limit of $a t \ll c$  
the radius of curvature is $R_c \simeq  c^2 / (a \sin \theta)$,
where $\theta$ is the angle between the initial direction of the 
 field line 
and the acceleration (see  figure 2  by Singal$^{(5)}$, 1997).  
 The force per unit volume due  to   the electric stress is

$$   f_E (r) = {{E^2}(r)\over{4 \pi R_c}} = 
 {{a \sin \theta}\over{c^2}} {{e^2}\over{4 \pi r^4}} , \eqno(4)  $$   

\no where in the second equality we have substituted for the electric 
field $E=e/r^2$. 
 The stress force is perpendicular to the direction of the field lines, 
so that the component of the stress force along the acceleration is
$ - f_E(r) \sin \phi$, where  $\phi$ is the angle between the 
local field line and the acceleration.   For very short intervals (where 
the direction of the  field lines did not change much 
 from their original  direction) $\phi \sim \theta$, 
and we can write: 
$ - f_E(r) \sin \phi \simeq - f_E(r) \sin \theta  = 
  {-a \sin^2 \theta \over c^2} {e^2 \over 4 \pi r^4}$.   
The dependence of this force on $\theta$ is similar to the dependence of the radiation 
distribution of an  accelerated charge at zero velocity on $\theta$.      
Integration of this force  over a spherical shell extending from $r=c \Delta t$ 
to $r_ {\rm up}$, where $c^2 / a \gg r_{\rm up} \gg c \Delta t$,
yields the total force due to stress 

$$  F_E(t) = 2 \pi  \int _{c \Delta t}^{r_{\rm up}} r^2 dr 
\int_0^{\pi} \sin \theta d \theta
[- f_E(r) \sin \theta ]
= - {{2}\over{3}} {{a}\over{c^2}} {{e^2}\over{c \Delta t}}
\left(1-{{c \Delta t}\over{r_{\rm up}}} \right) . \eqno(5)  $$
 
 Clearly the second term in the parenthesis can be neglected.
(Preliminary analysis done by us shows that, with certain limitations, 
nonconstant acceleration can be incorporated into 
our picture. This will be presented elsewhere.)
The power supplied by the external force on acting against the
electric stress is $P_E= - F_E  v = - F_E a \Delta t $, 
where we substituted  $v= a \Delta t$, and  $v$  is the charge velocity 
 in the system $S$, defined earlier,  at time $t = \Delta t$.   
Substituting for  $F_E$   we obtain 

$$ P_E(t) = {{2}\over{3}} {{a^2 e^2}\over{c^3}} . \eqno(6)   $$  

 This is the power radiated by the accelerated charged particle (eq. 3). 

\bg
\no{\bf  5.  Conclusions }
\bg

 As stated earlier, the cause of the radiation of 
an accelerated charge is the electric stress created when the field 
lines of the charge are curved relative to the system of reference adjacent 
to the charge.  The rate of work required to overcome the 
electric stress of the charged particle is equal to the power of the 
radiation emitted by the accelerated charge.

We already mentioned that for a zero velocity,  the radiation of 
a charge accelerated  
at a uniform acceleration,  $a$,  
 imparts no momentum to the radiating charge.    
This is the 
cause  why a radiation formed by a uniform acceleration does not 
create a radiation reaction force.   Intuitively, one tends to 
consider the radiation reaction force as the force that contradicts 
the creation of radiation, and the work done through overcoming 
this force, is the work invested in creating the radiation.   However, 
the fact that radiation formed by a uniform acceleration does not 
create a reaction force, raised difficulties about the energy balance 
of the system, and people tried to explain this difficulty 
in different ways.   
  (e. g. Leibovitz \& 
Peres$^{(7)}$  1963, Parrot $^{(9)}$ 1997).   
The existence of the stress force of the curved electric field, 
 supplies  a simple explanation for the source of the energy of the 
radiation, which is the work done by the force that overcomes the 
stress force of the curved electric field.   
Thus the paradox of the energy balance in radiation of a uniformly 
accelerated charge is solved.   

We would like to stress here again that the calculations  carried 
by Rohrlich$^{(1)}$ (1965)  about the relative acceleration 
are correct, but we find that we should consider the relative acceleration 
between the charge and its (curved) electric field, rather than   
 the relative acceleration between the charge and the observer.

\bg

\no {\bf Acknowledgment:}
 During early stages of this project we benefited 
from our discussions with  the late Nathan Rosen.  We 
acknowledge useful discussions on this topic with Amos Ori,  
 Hillel Livne, and Ofer Eyal.

We also acknowledge the comments of the referee that helped 
to improve the manuscript.

\vfil
\eject 
                                                         
\no {\bf references}
\bg

 \no (1)  Rohrlich, F. 1965, in {\it Classical Charged Particles}, Addison-Wesley Pub. Co. 
     
 \no (2) Rohrlich, F. 1963, Annals of Physics, 22, 169.
 
\no (3) Coleman, S.  1961,  Project Rand,  RM- 2820-PR.   
 
 \no (4)  Boulware, D. G. 1980, Annals of Physics, 124, 169.

 \no (5)   Singal, A. K. 1997, Gen. Rel. Grav., 29, 1371.   

 \no (6)  Rindler, W. 1966,  {\it Special Relativity}, Second Edition,
Oliver and Boyd.   

 \no (7)  Leibovitz, C. \& Peres, A.  1963,  Annals of Physics, 
 25, 400.

 \no (8)  Jackson, J. D. 1975, {\it Classical Electrodynamics}, 
Second Edition, John Wiley \& Sons (New York).

 \no (9)   Parrot, S.  1997, Gen. Rel. Grav., 29, 1463.  
 
 \no (10)  Rosen, N. 1962, Annals of Physics,  17, 269.  
 
 \no (11)   Singal, A. K. 1995, Gen. Rel. Grav., 27, 953.

\end{document}